\documentclass[twocolumn,aps,prb,showpacs,amsmath,superscriptaddress]{revtex4-1}
\usepackage{amssymb}
\usepackage{amsmath}
\usepackage{amsfonts}
\usepackage{CJK}
\usepackage{color}
\usepackage{epsfig}
\usepackage{float}
\usepackage{graphicx}
\usepackage{mathrsfs}
\usepackage{mathrsfs}
\usepackage{natbib}
\usepackage[toc,page,title,titletoc,header]{appendix}
\usepackage{subfigure}
\usepackage{ulem}
\usepackage{bm}

\begin{document}

\title{Characterization of Lifshitz transitions in topological nodal line semimetals}
\author{Hui Jiang}
\affiliation{Beijing National Laboratory for Condensed
Matter Physics, Institute of Physics, Chinese Academy of Sciences, Beijing 100190,
China}
\affiliation{School of Physical Sciences, University of Chinese Academy of Sciences, Beijing, 100049, China}
\author{Linhu Li}
\email{phylli@nus.edu.sg}
\affiliation{Department of Physics, National University of Singapore, 117542, Singapore}
\author{Jiangbin Gong}
\affiliation{Department of Physics, National University of Singapore, 117542, Singapore}
\author{Shu Chen}
\email{schen@iphy.ac.cn}
\affiliation{Beijing National Laboratory for Condensed
Matter Physics, Institute of Physics, Chinese Academy of Sciences, Beijing 100190,
China}
\affiliation{School of Physical Sciences, University of Chinese Academy of Sciences, Beijing, 100049, China}
\affiliation{Collaborative Innovation Center of Quantum Matter, Beijing, China}
\begin{abstract}
We introduce a two-band model of three-dimensional nodal line semimetals, the Fermi surface of which at half-filling may form various one-dimensional configurations of different topology. We study the symmetries and ``drumhead" surface states of the model, and find that the transitions between different configurations, namely, the Lifshitz transitions, can be identified solely by the number of gap-closing points on some high-symmetry planes in the Brillouin zone. A global phase diagram of this model is also obtained accordingly. We then investigate the effect of some extra terms analogous to a two-dimensional Rashba-type spin-orbit coupling. The introduced extra terms open a gap for the nodal line semimetals and can be useful in engineering different topological insulating phases.  We demonstrate that the behavior of surface Dirac cones in the resulting insulating system has a clear correspondence with the different configurations of the original nodal lines in the absence of the gap terms.
\end{abstract}

\date{\today}
\maketitle

\section{Introduction}
\indent
Exploring novel topological phases has been one of the most fruitful avenues in condensed matter physics during the past decade. As a new type of topological phases of matter, nodal line semimetals (NLSMs) have attracted great attention in both theoretical \cite{Burkov2011,Weng2015,Kim2015,Yu2015,Zhang2016,Yan2016,Lim2017,Li2017,Li2017_2}  and experimental \cite{Wu2016,Bian2016,Hu2016,Yan2017} studies during the past years. Compared with the well-known Dirac and Weyl semimetals \cite{Wan2011,Young2012,Morimoto2014,Yang2014} characterized by discrete zero-dimensional (0D) band-crossing points in momentum space, NLSMs have one-dimensional (1D) band-crossing lines of the conduction and valence bands, which possess much richer configurations of Fermi surface, e.g. from the simplest case with a single loop, to the more complicate linked loops\cite{Zhong2017,Chen_Hopf,Yan_Hopf,Ezawa_Hopf,Chang_Hopf}, knots\cite{Ezawa_Hopf,Bi2017} and crossing lines\cite{Bzdusek2016,Yan_crossing,Ezawa_crossing,Yan2017}.

 One topological invariant characterizing a nodal line of NLSMs is the Berry phase, defined along a trajectory either enclosing a 1D band-crossing line, or with one momentum varying over a period while the others being fixed \cite{Burkov2011,Zhang2016,Li2017_2}. However, this invariant alone does not suffice to describe NLSMs because it only unveils the local topological properties in momentum space, whereas the topology of the whole Fermi surface is related to the global information of the whole Brillouin zone (BZ).  
Because the gap-closing lines are in general related to certain symmetries, NLSMs may be classified by the behavior of the lines on a symmetry-related plane in the BZ \cite{Okugawa2017}. More generally, different configurations of nodal lines are associated with the topology of Fermi surface, and the transitions between them can be understood as the Lifshitz transitions \cite{Lifshitz}, which have been considered for NLSMs very recently \cite{Lifshitz_NL}. This way, a sudden change of the topology of the Fermi surface between a variety of topologically inequivalent configurations can be captured by the Lifshitz transitions.


In some recent studies, NLSMs have been taken as bases from which numerous topological phases can be generated by adding some extra terms, e.g. a 2D Chern insulator \cite{Li2016}, a 3D Chiral insulator \cite{Li2017}, a nodal torus semimetal \cite{Turker2017} and a 2nd order topological insulator \cite{Li2017_3}.
Moreover, it has been shown that the topological properties of the resulting system can be characterized by the behavior of extra terms along the nodal lines in the original NLSM, which means the nodal lines can serve as indicators of topological properties of the system. Inspired by this fact, an interesting question to ask is whether we can do it the other way around, i.e. to distinguish between different configurations of nodal lines by considering extra terms added to the system, and observing the consequences?

In this paper, we consider a simple 2-band NLSM, which exhibits Lifshitz transitions between several different configurations of the Fermi surface. The gap-closing lines are protected by the time-reversal symmetry and inversion symmetry together, while extra mirror symmetries in this model ensure that the surface states can only be observed under open boundary conditions (OBCs) in specific directions. We then show that the geometric structure of these gapless lines can be determined by only looking at the high-symmetry planes in momentum space, as the number of gap-closing points on the planes of $k_z=0$ or $\pi$ can exactly distinguish different Lifshitz transition regimes. Finally, we induce extra terms analogous to a 2D Rashba-type spin-orbit coupling to our model, and find that the different configurations of nodal lines can be manifested by the behavior of surface Dirac cones in the resulting system.

\section{Model Hamiltonian and the geometry of lines} \label{a}
To model a NLSM, it is necessary for the system to have at least two bands, thus the Hamiltonian of which can be written as
\begin{eqnarray}
H(\bm{k})=\bm{h}(\bm{k})\cdot\bm{\sigma},
\end{eqnarray}
with $\bm{\sigma}$ the Pauli matrices acting on a pseudospin-1/2 (e.g. orbitals) space. Generally speaking, the absence of one Pauli matrix guarantees the Hamiltonian to have a gapless 1D region in its spectrum. In this work we consider a NLSM given by
 \begin{equation}\label{eq1.13a}
H_{\mathrm{NL}}(\bm{k})=(\cos k_x+\cos k_y-m)\sigma_x+(\cos k_z+b \cos k_y)\sigma_z,
\end{equation}
with $m$ and $b$ the parameters to control the configuration of the nodal lines.
This model satisfies both the time-reversal symmetry represented by a complex conjugation operation for spinless systems, and an inversion symmetry given by $H_{\mathrm{NL}}(\bm{k})=H_{\mathrm{NL}}(\bm{-k})$. The combining $PT$ symmetry $H_{\mathrm{NL}}^*(\bm{k})=H_{\mathrm{NL}}(\bm{k})$ ensures the absence of $\sigma_y$ and the existence of gapless nodal lines. Furthermore, this model also satisfies mirror symmetries in each of the three directions, i.e. $H_{\mathrm{NL}}(k_i)=H_{\mathrm{NL}}(-k_i)$ with $i=x,y,z$. These symmetries give some restrictions to the behavior of the nodal lines, and make it possible to characterize the Lifshitz transitions by only looking at some high-symmetry planes, as discussed later in the paper.

\begin{figure}[ht]
\setlength{\abovecaptionskip}{0pt}
\setlength{\belowcaptionskip}{10pt}
\includegraphics[width=1\linewidth]{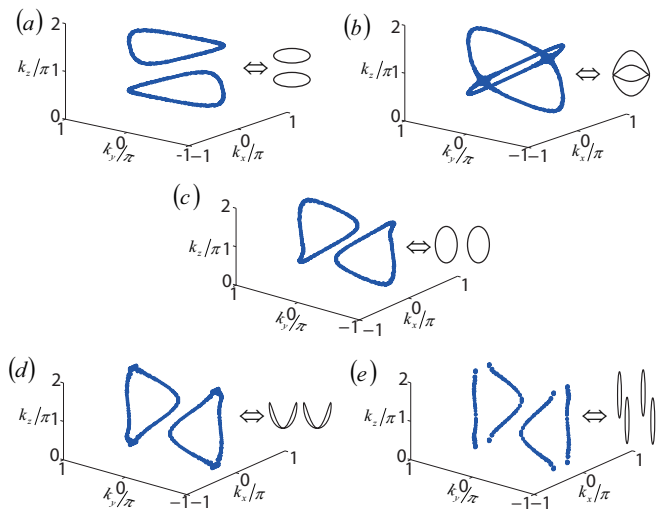}
\caption{The geometry of the Fermi surface with $m=0.5$. The values of $b$ in
(a) ,(b), (c), (d) and (e)  are $b=0.5$, $1$, $1.5$, $2$, $2.5$, respectively. The insets show the graphs geometrically equivalent to corresponding gapless lines. Results are obtained numerically, by considering the regime where $E_{\mathrm{NL}}$ in Eq. (\ref{eigen}) is less than $10^{-3}$.} \label{fig1}
\end{figure}

The energy dispersion of the model is given by
\begin{eqnarray}
E_{\mathrm{NL}}=\pm\sqrt{B_x^2+B_z^2},\label{eigen}
\end{eqnarray}
and the nodal lines consist of points where $E_{\mathrm{NL}}=0$, the conditions of which are given by:
\begin{eqnarray}
B_x &=& \cos k_x+\cos k_y-m=0, \label{condition1}\\
B_z &=& \cos k_z+b \cos k_y=0.\label{condition2}
\end{eqnarray}
Depending on the value of $m$ and $b$, the system supports several different configurations of gap-closing lines, as illustrated in Fig. \ref{fig1}. Without loss of generality, we consider only positive $m$ and $b$ throughout the paper, and here we fixed $m=0.5$ as an example to demonstrate the different types of nodal lines. In this case Eq. (\ref{condition1}) gives a cylinder along $k_z$ in the BZ, while Eq. (\ref{condition2}) gives two planes parallel to $k_x$ axis, and the 1D nodal lines are given by the intersection of the two 2D manifolds. When $0\leq b<1$,  this system holds two loops which can be mapped to each other by the mirror symmetry in $k_z$ direction, as shown in Fig. \ref{fig1}(a). By increasing $b$, the two loops deform and touch each other when $b=1$ [Fig. \ref{fig1}(b)], and transform into two  loops mirror-symmetric with respect to the $k_y=0$ (or $\pi$) plane when $1<b<2$ [Fig. \ref{fig1}(c)]. Further increasing $b$, these two loops will extend along $k_z$ and merge with themselves at the boundary of the periodic BZ when $b=2$, as shown in [Fig. \ref{fig1}(d)]. When $b>2$, the two loops divide into four lines going through the BZ, which can be taken as four closed loops due to the periodic condition of the BZ, as shown in [Fig. \ref{fig1}(e)]. In each panel we inset a simpler sketch which is topologically equivalent to the corresponding nodal lines. The Lifshitz transitions take place at $b=1$ and $b=2$, where the deformations of configurations are not continuous (gluing with each other or theirselves). The configurations of Fermi surface at these transition points correspond to nodal chain semimetals \cite{Bzdusek2016,Yan2017}.

\section{Surface states and Phase Diagram}
It is known that a topological NLSM has ``drumhead" surface states within the regime enclosed by the projection of the nodal lines onto a 2D surface BZ. These surface states are associated with a nontrivial topological invariant of the bulk states, i.e. a Berry phase
\begin{eqnarray}
\gamma_c=\oint_c \langle \psi | i \partial_{\theta}|\psi\rangle
\end{eqnarray}
along a trajectory $c$ enclosing a nodal line, with $\psi$ the occupied Bloch band and $\theta$ the phase angle of the trajectory $c$. Alternatively, the trajectory $c$ can be continuously deformed into two lines $l_1$ and $l_2$ passing through the BZ with two momenta being fixed, and the Berry phases $\gamma_{l_{1,2}}$ along these two lines satisfies \cite{Zhang2016}
\begin{eqnarray}
\gamma_c=\gamma_{l_1}-\gamma_{l_2}~\mathrm{mod}~2\pi.
\end{eqnarray}
Since each $\gamma_l$ is defined with two fixed momenta, i.e. fixed $\bm{k}_{\parallel}=(k_i,k_j)$ with $i,j$ taking two of $x$, $y$ and $z$, it describes the topology along a quasi-1D system along the third momentum $k_{\perp}$ direction for a fixed point in the 2D BZ of $\bm{k}_{\parallel}$. The existence of surface states suggests a $\pi$ Berry phase, which is equivalent to one of the two $\gamma_l$ given by $\pi$. 

In our model, due to the existence of the mirror symmetries, a given line $l$ with fixed $\bm{k}_{\parallel}$ is enclosed by either both of the two loops (or the two pairs of the four lines when $b>2$) or none of them, with $\bm{k}_{\parallel}$ contains any two of $x$, $y$ and $z$. Therefore, for two straight lines $l_1$ and $l_2$ at $\bm{k}_{\parallel,1}$ and $\bm{k}_{\parallel,2}$, they can continuously deform into either a point and disappear if they are at the same side of the two loops (or the two pairs of the four lines), or two trajectories $c_1$ and $c_2$ enclosing each of the two loops (or the two pair of the four lines) if they are at the different sides of them, which leads to
\begin{eqnarray}
\gamma_{l_1}-\gamma_{l_2}=\gamma_{c_1}+\gamma_{c_2}=0~\mathrm{mod}~2\pi.
\end{eqnarray}
In either case, we can see that there is no topological difference between the quasi-1D systems at $\bm{k}_{\parallel,1}$ and $\bm{k}_{\parallel,2}$.
Hence the nontrivial topologies of the two nodal loops (or the two pairs of the four lines) always cancel out each other and make no difference when considering the existence of surface states at a given $\bm{k}_{\parallel}$. Consistent with this symmetry-based perspective, numerical results have shown no surface state when OBC is taken in $x$, $y$ or $z$ direction. We would also like to point out that in some cases, a system with Berry phase $\gamma_{l}=0~(\mathrm{mod}~2\pi)$ may also have four-fold surface states, which correspond to a winding number of $2$ along $l$ \cite{Li2017_2}. In our model, however, the Berry phases $\gamma_{c_{1,2}}$ of the two loops (or the two pairs of the four lines) always correspond to opposite windings of the pseudospin texture, hence a straight line $l$ enclosed by both of them can only have a total winding of zero, and as such this still suggests no surface states.

\begin{figure}[ht]
\setlength{\abovecaptionskip}{0pt}
\setlength{\belowcaptionskip}{10pt}
\includegraphics[width=1\linewidth]{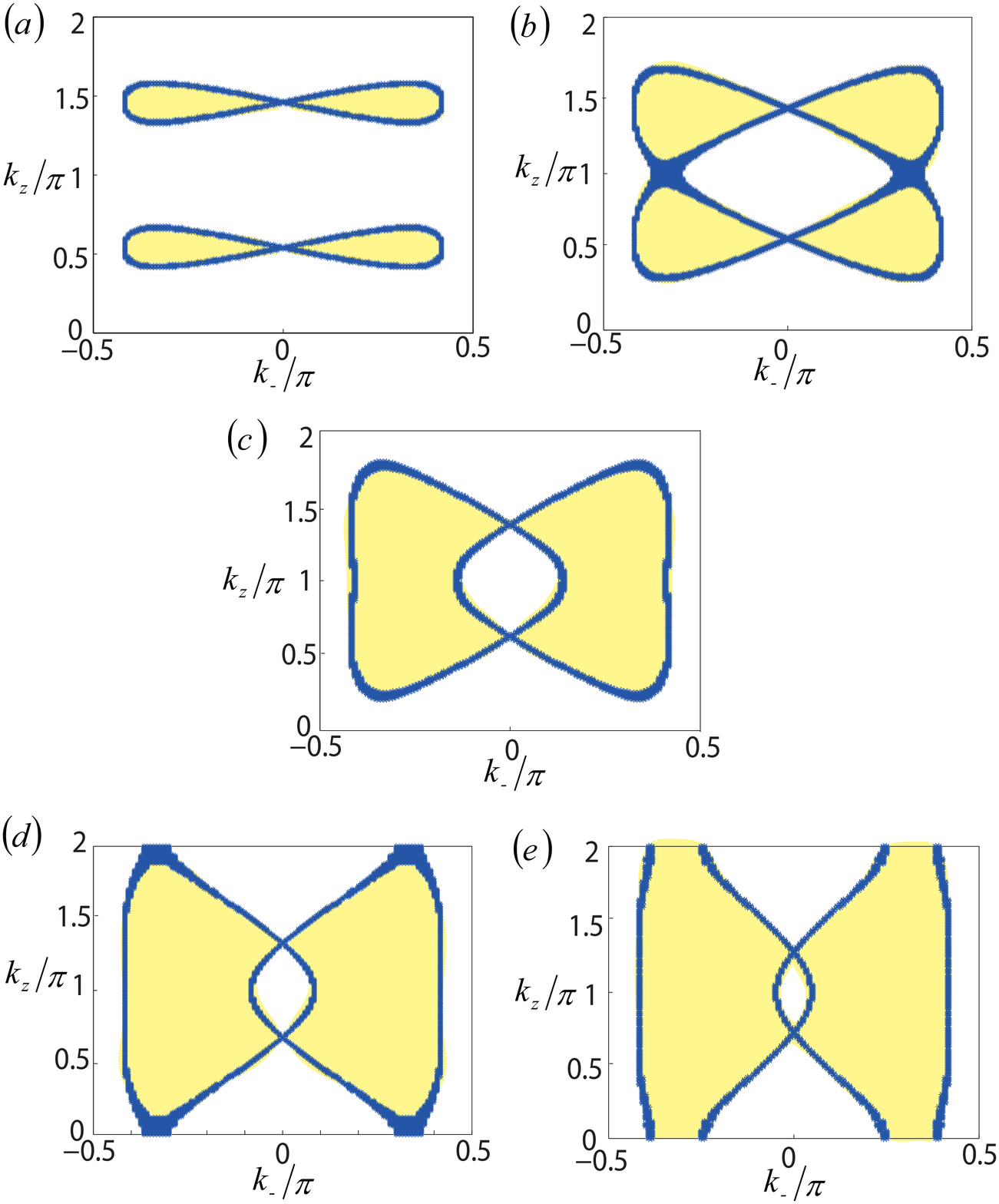}
\caption{``Drumhead" surface states with OBC along $k_+$ direction. The projections of the nodal lines are shown by the blue lines, which are obtained from the same method as in Fig. \ref{fig1}. The ``Drumhead" surface states exist in the yellow (shadow) areas. The figures are with the same parameters as in Fig. \ref{fig1}, i.e. $m=0.5$ and $b=0.5$, $1$, $1.5$, $2$ and $2.5$ from (a) to (e).} \label{fig2}
\end{figure}

Nevertheless, the nontrivial topology of the nodal lines can be manifested by surface states when OBC is taken in a direction without mirror symmetry. Specifically, we consider a rotated axis with $k_{\pm}=\frac{1}{2}(k_x\pm k_y)$, as the model does not possess mirror symmetry along each of $k_{\pm}$ directions. As a consequence, ``drumhead" surface states emerge when OBC is taken in either $k_+$ or $k_-$ direction. In order to illustrate the ``drumhead" states, we numerically diagonalize the Hamiltonian with OBC along $k_+$ (i.e. $x+y$) direction, while $k_z$ and $k_-$ are taken as fixed parameters. In Fig. \ref{fig2} we show the areas [marked in yellow (shadow)] with degenerate zero-energy surface states, which are enclosed by the projections of nodal lines [blue (dark) lines] in the 2D surface BZ of $k_z$ and $k_-$.

\begin{figure*}
\setlength{\abovecaptionskip}{0pt}
\setlength{\belowcaptionskip}{10pt}
\includegraphics[width=0.8\linewidth]{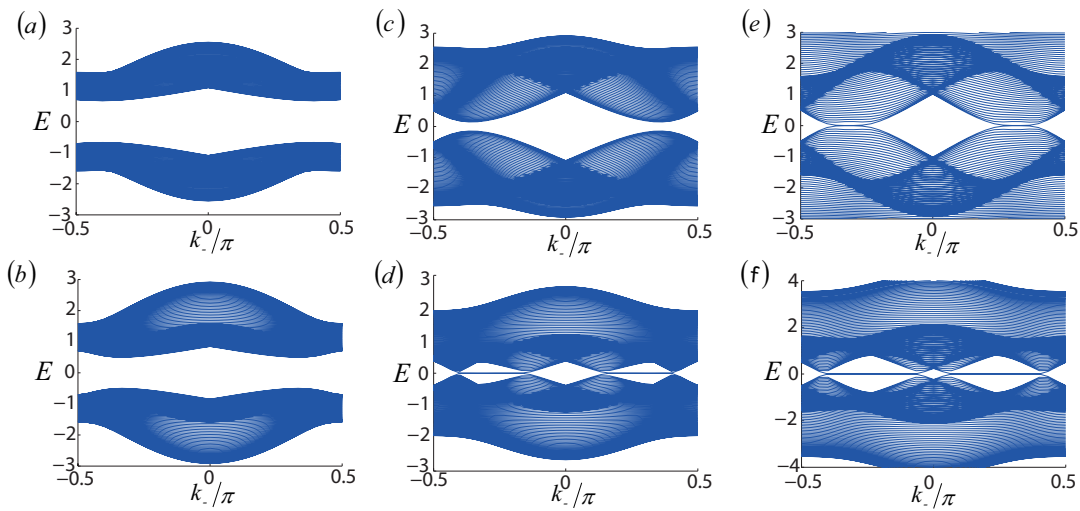}
\caption{ The spectra as a function of $k_-$. $k_z$ is fixed at $0$ or $\pi$, and OBC is taken along $k_+$ direction with $100$ lattice sites.
The parameters are $m=0.5$ and $b=0.5$,$1.5$ and $2.5$ from left to right.} \label{fig3}
\end{figure*}

Next we focus on the high-symmetry planes of $k_z=0$ and $\pi$, as the mirror symmetry along $k_z$ is still preserved in the new coordinates. We find that the non-continuous deformation in our model only occurs at these planes, therefore we can characterize the Lifshitz transitions by focusing on the gapless points and surface states with $k_z$ fixed at these two values. In Fig. \ref{fig3} we demonstrate the spectra as a function of $k_-$, with OBC along $k_+$ direction and $k_z=0$ and $\pi$. The different configurations in Fig. \ref{fig1} clearly show different features of surface states on the high-symmetry planes. When the two loops are connected by mirror symmetry along $k_z$ [Fig. \ref{fig1}(a)], there is no gapless point or surface state in either of the two planes, as shown in Fig. \ref{fig3}(a) and (b). When $1<b<2$, the two loops are connected by mirror symmetry along $k_y$[Fig. \ref{fig1}(c)],  and four gapless points with surface states connecting them emerge only for $k_z=\pi$ as shown in Fig. \ref{fig3}(c) and (d). Finally, when the two loops divide into four lines [Fig. \ref{fig1}(e)], gapless points and surface states also emerge for $k_z=0$, as shown in Fig. \ref{fig3}(e) and (f).

\begin{figure}[ht]
\setlength{\abovecaptionskip}{0pt}
\setlength{\belowcaptionskip}{10pt}
\includegraphics[width=1\linewidth]{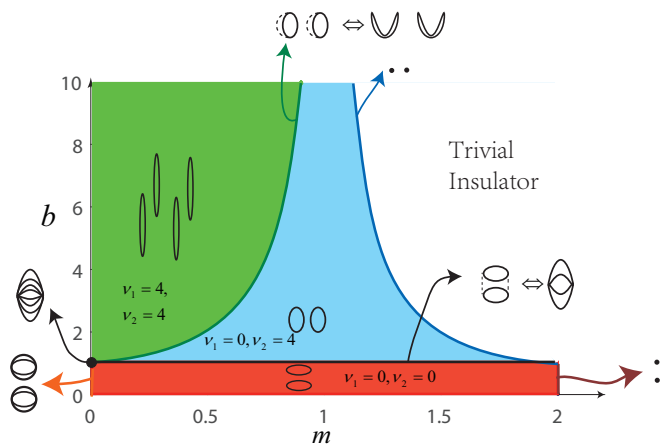}
\caption{Phase diagram of Hamiltonian (\ref{eq1.13a}) versus $m$ and $b$, insets are graphs geometrically equivalent to corresponding gapless regimes in the BZ. $v_1$ and $v_2$ are the number of gapless points on planes of $k_z=0$ and $=\pi$ respectively.} \label{fig4}
\end{figure}

Lastly, we use this characterization to establish a phase diagram of our model, as shown in Fig. \ref{fig4}. As the non-continuous deformations are characterized by gluing and tearing the loops at the two high-symmetry planes, we can simply count the number of gap-closing points, i.e. the solutions of Eqs. (\ref{condition1}) and (\ref{condition2}), at $k_z=0$ and $\pi$ to distinguish between different phases.
Namely, the system is i) a trivial insulator when $b< 1,m>2$ or $b>\frac{1}{m-1}>0$; ii) a NLSM with two nodal loops of different orientations in the regimes given by $b<|\frac{1}{m-1}|$ and $b< 1,m<2$ respectively; and iii) a NLSM with four nodal lines through the BZ when $b >\frac{1}{1-m}>0$. Sketches of the configurations of gapless regions in the BZ are also shown in the diagram. Along the phase boundaries, there are several configurations which have not been discussed before. At the transition boundary from a semimetal to an insulator, the nodal loops shrink into two points and disappear when increasing $m$. When $m=0,b<2$, each of the two loops extends to the edge of the BZ and connects with itself at two points. These two self-connected loops touch each other when $m=0,b=2$, and divide into four lines when $m=0,b>2$. We also point out that although this analysis is restricted in the regime with non-negative parameters $m$ and $b$, our results can be extended to arbitrary $m$ and $b$ by considering certain transformations of the Hamiltonian. Explicitly, the Hamiltonian is invariant under the operations $C_1:~\sigma_yH(k_x,k_y,k_z,m,b)\sigma_y=H(k_x+\pi,k_y+\pi,k_z+\pi,-m,b)$, and $C_2:~\sigma_xH^*(k_x,k_y,k_z,m,b)\sigma_x=H(k_x,k_y,k_z+\pi,m,-b)$. These two operations map the positive $m$ and $b$ to negative ones respectively, and the combination of $C_1$ and $C_2$ reverses signs of both the two parameters.

\section{effect of extra gap terms}
\begin{figure}
\includegraphics[width=0.8\linewidth]{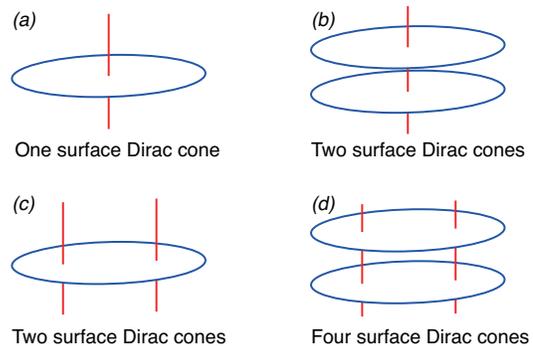}
\caption{Sketches of different geometric relations between zero-lines or extra terms (red lines), and the nodal loops from the original Hamiltonian (blue circles). The number of corresponding surface Dirac cones are indicated under the figures.} \label{fig5}
\end{figure}

It has been shown that the a NLSM can be turned to an insulator by introducing some extra gap terms, and the behavior of surface states of the resulting insulator has a correspondence with the original gap-closing lines of the NLSM \cite{Li2016,Li2017,Li2017_3}. In other words, one can obtain the information of Fermi surface by adding some extra physical effects to open a band gap, and investigating the surface states of the resulting insulator. More precisely, suppose the zeros of extra terms form one or several 1D lines (referred as zero-lines here after), the surface states shall have a Dirac cone whenever such a line is enclosed by a gap-closing loop in the original NLSM \cite{Li2017}. In Fig. \ref{fig5} we show several examples of geometric relations between zero-lines and the original nodal loops, which correspond to different number of surface Dirac cones.

In order to distinguish between different configurations of nodal lines in our model, the zero-lines of extra terms need to have different geometric relations with each configuration. Here we consider the following extra terms added to the model:
\begin{eqnarray}
H_{\mathrm{gap}}&=&\sin{k_z}\tau_x\sigma_y+\sin{k_x}\tau_y\sigma_y,\label{gap}
\end{eqnarray}
with $\tau_i$ the Pauli matrices acting on another (pseudo)spin-1/2 space. These gap terms can be viewed as a 2D Rashba type (pseudo)spin-orbit coupling along $k_y$ direction, which also couples the two components of $\sigma$. The zeros of these terms form four straight zero-lines through the BZ along $k_y$, with $k_x$ and $k_z$ equal to either $0$ or $\pi$. Due to the anticommuting relation between $H_{\mathrm{gap}}$ and the Hamiltonian (\ref{eq1.13a}), the eigen-energies of the resulting system are doubly degenerate, and take the form of
\begin{eqnarray}
E=\pm\sqrt{E_{\mathrm{NL}}+E_{\mathrm{gap}}},
\end{eqnarray}
with $E_{\mathrm{gap}}=\pm\sqrt{\sin^2{k_y}+\sin^2{k_z}}$.

\begin{figure*}
\includegraphics[width=1\linewidth]{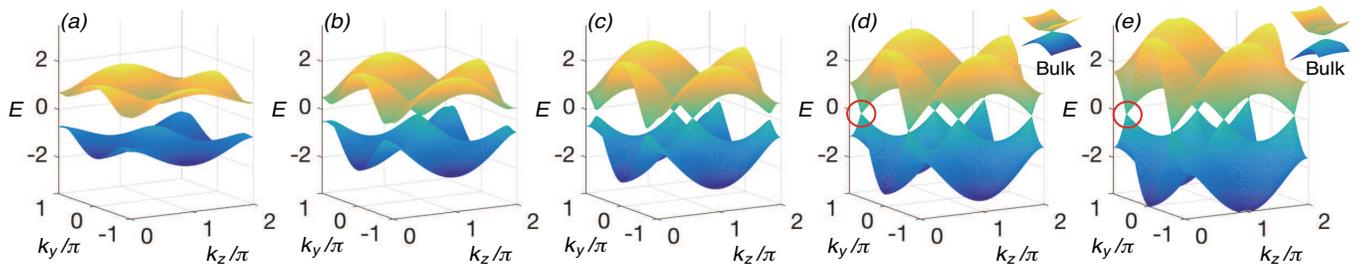}
\caption{The spectra of the two doublet bands nearest to $E=0$ for the total Hamiltonian $H=H_{\mathrm{NL}}+H_{\mathrm{gap}}$. The parameters are $m=0.5$, and $b=0.5,1,1.5,2,2.5$ from (a) to (e). The insets in (d) and (e) show the next two doublet bands nearest to $E=0$ around the region of the red circles, which correspond to bulk states. The bulk gap-closes at $b=2$ as shown by the inset of (d).} \label{fig6}
\end{figure*}

In Fig. \ref{fig6} we illustrate the two doublet bands nearest to $E=0$ under OBC along $x$, with the same parameters as in Fig. \ref{fig1}. The different configurations of gap-closing lines in the original NLSM can be directly distinguished by the surface Dirac cones in the spectrum. We note that with the parameter we choose, the zero-lines never cross the nodal lines except for $m=0.5,b=2$ in Fig. \ref{fig6}(d), hence the system is an insulator in the bulk with the parameters in Fig. \ref{fig6}(a)-(c) and (e). Therefore the gapless dispersion in corresponding panels must exist on the surface of the system.

When $b<1$, the two nodal loops do not enclose any of the zero-lines, and there is no any gapless states in the spectrum [Fig. \ref{fig6}(a)]. When $b=1$, the two loops deform into a pair of crossing loops which encloses the zero-line at $k_z=k_x=0$, and surface gapless states with quadratic dispersion emerge at $k_y=k_z=0$ when OBC is taken along $x$ [Fig. \ref{fig6}(b)]. In the regime with $1<b<2$, the crossing loops divide into two separate loops, and the quadratic states also divide into two surface Dirac cones in Fig. \ref{fig6}(c), as a zero-line is enclosed by both loops. Keep increasing $b$, we can see that the bulk gap closes at $b=2$ [Fig. \ref{fig6}(d)] and reopens after this point, and leave another two surface Dirac cones [Fig. \ref{fig6}(e)]. This pair of Dirac cones corresponds to the zero-line at $k_x=0$ and $k_z=\pi$, which now is also enclosed by the two ``effective loops", each of them constructed by a pair of the four nodal lines in Fig. \ref{fig1}(e).

However, we need to point out that it is not sufficient to distinguish different configurations of the original nodal lines by only looking at the number of surface Dirac cones in a system with fixed parameters. For example, the two different cases in Fig. \ref{fig5}(b) and (c) both have two surface Dirac cones. To see the difference, one need to observe how the surface Dirac cones behave when the parameters are tuned continuously.  By fixing the extra terms and tuning the parameters in the nodal line Hamiltonian (\ref{eq1.13a}), the two nodal loops in Fig. \ref{fig5}(b) can merge and reshape into two loops, none of them encloses the zero-line. In this procedure the nodal loops never cross the zero-line, hence the bulk gap is always opened, and one shall observe the two Dirac cones merge into one quadratic gapless point and annihilate, which is the case from Fig. \ref{fig6}(c) to (a). Whereas in Fig. \ref{fig5}(c), the two Dirac cones are related to two zero-lines within the same loop, hence they cannot disappear continuously by tuning only the nodal loop enclosing them, unless the nodal loop crosses the zero-lines and the system become gapless at the crossing point. On the other hand, if we fix the parameters in the nodal line Hamiltonian and tuning the extra terms, the two zero-lines in Fig. \ref{fig5}(c) may merge and disappear, thus their two corresponding surface Dirac cones may merge and annihilate with each other, while the ones related to Fig. \ref{fig5}(b) can not disappear in this way. 

\section{Summary} \label{e}

In summary, we have studied a simple 2-band model of NLSMs with $PT$ symmetry, which exhibits various configurations of nodal lines in momentum space, including two loops related to each other by mirror symmetries along $k_z$ or $k_y$ directions, and four nodal lines extending through the BZ. The transitions between these configurations, namely the Lifshiz transitions, correspond to non-continuous transformations of the nodal lines, which occur only at the high-symmetry planes with $k_z=0$ and $\pi$ in the BZ.
The existence of ``drumhead" surfaces of the NLSM, on the other hand, requires OBC in a direction where the system is not mirror-symmetric to itself, and their behavior in the two high-symmetry planes also reflects the Lifshitz transitions in our model. According to these features, we obtain a phase diagram regarding different configurations of nodal lines.
Finally, we introduce some extra terms analogous to a 2D Rashba-type spin-orbit coupling, which open a gap in the system. The behavior of surface Dirac cones in the resulting insulating system under OBCs can manifest the nodal lines in the original semimetallic system.

Finally, we would also like to point out that although it may not be an easy task to find a realization of real materials, it is possible to simulate our model and its Lifshitz transition using some artificial systems. One promising experimental setup is the circuit systems, which has been used to realizing different topological phases in recent years \cite{circuit1,circuit2,circuit3,circuit4}. The topological surface states can be detected by the topological boundary resonances appearing in the impedance read-out of the corresponding circuit \cite{circuit3}. Such realization requires the model to be time-reversal invariant, hence the extra terms of Eq. (\ref{gap}) in our model cannot be simulated directly. Nevertheless, the key feature here is to have two extra terms anticommuting with each other and with the nodal line Hamiltonian, therefore we can choose an alternative form as
\begin{eqnarray}
H_{\mathrm{gap},2}&=&\sin{k_z}\tau_x\sigma_y+\sin{k_x}\tau_z\sigma_y,
\end{eqnarray}
which obeys the time-reversal symmetry as $H^{*}_{\mathrm{gap},2}(\bm{k})=H_{\mathrm{gap},2}(-\bm{k})$, and the behavior of surface Dirac cones is exactly the same as discussed previously.

\begin{acknowledgments}
The work is supported by the National Key Research and Development Program of China (2016YFA0300600), NSFC under Grants No. 11425419, No. 11374354 and No. 11174360, and the Strategic Priority Research Program (B) of the Chinese Academy of Sciences  (No. XDB07020000). J.G. is supported by the Singapore NRF grant No. NRF-NRFI2017-04 (WBS No. R-144-000- 378-281) and by the Singapore Ministry of Education Academic Research Fund Tier I (WBS No. R-144-000-353-112).
\end{acknowledgments}

\bibliographystyle{apsrev4-1}
\bibliography{bibfile_loop}

\begin{thebibliography}{36}%
\makeatletter
\providecommand \@ifxundefined [1]{%
 \@ifx{#1\undefined}
}%
\providecommand \@ifnum [1]{%
 \ifnum #1\expandafter \@firstoftwo
 \else \expandafter \@secondoftwo
 \fi
}%
\providecommand \@ifx [1]{%
 \ifx #1\expandafter \@firstoftwo
 \else \expandafter \@secondoftwo
 \fi
}%
\providecommand \natexlab [1]{#1}%
\providecommand \enquote  [1]{``#1''}%
\providecommand \bibnamefont  [1]{#1}%
\providecommand \bibfnamefont [1]{#1}%
\providecommand \citenamefont [1]{#1}%
\providecommand \href@noop [0]{\@secondoftwo}%
\providecommand \href [0]{\begingroup \@sanitize@url \@href}%
\providecommand \@href[1]{\@@startlink{#1}\@@href}%
\providecommand \@@href[1]{\endgroup#1\@@endlink}%
\providecommand \@sanitize@url [0]{\catcode `\\12\catcode `\$12\catcode
  `\&12\catcode `\#12\catcode `\^12\catcode `\_12\catcode `\%12\relax}%
\providecommand \@@startlink[1]{}%
\providecommand \@@endlink[0]{}%
\providecommand \url  [0]{\begingroup\@sanitize@url \@url }%
\providecommand \@url [1]{\endgroup\@href {#1}{\urlprefix }}%
\providecommand \urlprefix  [0]{URL }%
\providecommand \Eprint [0]{\href }%
\providecommand \doibase [0]{http://dx.doi.org/}%
\providecommand \selectlanguage [0]{\@gobble}%
\providecommand \bibinfo  [0]{\@secondoftwo}%
\providecommand \bibfield  [0]{\@secondoftwo}%
\providecommand \translation [1]{[#1]}%
\providecommand \BibitemOpen [0]{}%
\providecommand \bibitemStop [0]{}%
\providecommand \bibitemNoStop [0]{.\EOS\space}%
\providecommand \EOS [0]{\spacefactor3000\relax}%
\providecommand \BibitemShut  [1]{\csname bibitem#1\endcsname}%
\let\auto@bib@innerbib\@empty
\bibitem [{\citenamefont {Burkov}\ \emph {et~al.}(2011)\citenamefont {Burkov},
  \citenamefont {Hook},\ and\ \citenamefont {Balents}}]{Burkov2011}%
  \BibitemOpen
  \bibfield  {author} {\bibinfo {author} {\bibfnamefont {A.~A.}\ \bibnamefont
  {Burkov}}, \bibinfo {author} {\bibfnamefont {M.~D.}\ \bibnamefont {Hook}}, \
  and\ \bibinfo {author} {\bibfnamefont {L.}~\bibnamefont {Balents}},\
  }\href@noop {} {\bibfield  {journal} {\bibinfo  {journal} {Phys. Rev. B}\
  }\textbf {\bibinfo {volume} {84}},\ \bibinfo {pages} {235126} (\bibinfo
  {year} {2011})}\BibitemShut {NoStop}%
\bibitem [{\citenamefont {Weng}\ \emph {et~al.}(2015)\citenamefont {Weng},
  \citenamefont {Liang}, \citenamefont {Xu}, \citenamefont {Yu}, \citenamefont
  {Fang}, \citenamefont {Dai},\ and\ \citenamefont {Kawazoe}}]{Weng2015}%
  \BibitemOpen
  \bibfield  {author} {\bibinfo {author} {\bibfnamefont {H.}~\bibnamefont
  {Weng}}, \bibinfo {author} {\bibfnamefont {Y.}~\bibnamefont {Liang}},
  \bibinfo {author} {\bibfnamefont {Q.}~\bibnamefont {Xu}}, \bibinfo {author}
  {\bibfnamefont {R.}~\bibnamefont {Yu}}, \bibinfo {author} {\bibfnamefont
  {Z.}~\bibnamefont {Fang}}, \bibinfo {author} {\bibfnamefont {X.}~\bibnamefont
  {Dai}}, \ and\ \bibinfo {author} {\bibfnamefont {Y.}~\bibnamefont
  {Kawazoe}},\ }\href@noop {} {\bibfield  {journal} {\bibinfo  {journal} {Phys.
  Rev. B}\ }\textbf {\bibinfo {volume} {92}},\ \bibinfo {pages} {045108}
  (\bibinfo {year} {2015})}\BibitemShut {NoStop}%
\bibitem [{\citenamefont {Kim}\ \emph {et~al.}(2015)\citenamefont {Kim},
  \citenamefont {Wieder}, \citenamefont {Kane},\ and\ \citenamefont
  {Rappe}}]{Kim2015}%
  \BibitemOpen
  \bibfield  {author} {\bibinfo {author} {\bibfnamefont {Y.}~\bibnamefont
  {Kim}}, \bibinfo {author} {\bibfnamefont {B.~J.}\ \bibnamefont {Wieder}},
  \bibinfo {author} {\bibfnamefont {C.}~\bibnamefont {Kane}}, \ and\ \bibinfo
  {author} {\bibfnamefont {A.~M.}\ \bibnamefont {Rappe}},\ }\href@noop {}
  {\bibfield  {journal} {\bibinfo  {journal} {Phys. Rev. Lett.}\ }\textbf
  {\bibinfo {volume} {115}},\ \bibinfo {pages} {036806} (\bibinfo {year}
  {2015})}\BibitemShut {NoStop}%
\bibitem [{\citenamefont {Yu}\ \emph {et~al.}(2015)\citenamefont {Yu},
  \citenamefont {Weng}, \citenamefont {Fang}, \citenamefont {Dai},\ and\
  \citenamefont {Hu}}]{Yu2015}%
  \BibitemOpen
  \bibfield  {author} {\bibinfo {author} {\bibfnamefont {R.}~\bibnamefont
  {Yu}}, \bibinfo {author} {\bibfnamefont {H.}~\bibnamefont {Weng}}, \bibinfo
  {author} {\bibfnamefont {Z.}~\bibnamefont {Fang}}, \bibinfo {author}
  {\bibfnamefont {X.}~\bibnamefont {Dai}}, \ and\ \bibinfo {author}
  {\bibfnamefont {X.}~\bibnamefont {Hu}},\ }\href@noop {} {\bibfield  {journal}
  {\bibinfo  {journal} {Phys. Rev. Lett.}\ }\textbf {\bibinfo {volume} {115}},\
  \bibinfo {pages} {036807} (\bibinfo {year} {2015})}\BibitemShut {NoStop}%
\bibitem [{\citenamefont {Zhang}\ \emph {et~al.}(2016)\citenamefont {Zhang},
  \citenamefont {Zhao}, \citenamefont {Liu}, \citenamefont {Xue}, \citenamefont
  {Zhu},\ and\ \citenamefont {Wang}}]{Zhang2016}%
  \BibitemOpen
  \bibfield  {author} {\bibinfo {author} {\bibfnamefont {D.-W.}\ \bibnamefont
  {Zhang}}, \bibinfo {author} {\bibfnamefont {Y.~X.}\ \bibnamefont {Zhao}},
  \bibinfo {author} {\bibfnamefont {R.-B.}\ \bibnamefont {Liu}}, \bibinfo
  {author} {\bibfnamefont {Z.-Y.}\ \bibnamefont {Xue}}, \bibinfo {author}
  {\bibfnamefont {S.-L.}\ \bibnamefont {Zhu}}, \ and\ \bibinfo {author}
  {\bibfnamefont {Z.~D.}\ \bibnamefont {Wang}},\ }\href@noop {} {\bibfield
  {journal} {\bibinfo  {journal} {Phys. Rev. A}\ }\textbf {\bibinfo {volume}
  {93}},\ \bibinfo {pages} {043617} (\bibinfo {year} {2016})}\BibitemShut
  {NoStop}%
\bibitem [{\citenamefont {Yan}\ and\ \citenamefont {Wang}(2016)}]{Yan2016}%
  \BibitemOpen
  \bibfield  {author} {\bibinfo {author} {\bibfnamefont {Z.}~\bibnamefont
  {Yan}}\ and\ \bibinfo {author} {\bibfnamefont {Z.}~\bibnamefont {Wang}},\
  }\href@noop {} {\bibfield  {journal} {\bibinfo  {journal} {Phys. Rev. Lett.}\
  }\textbf {\bibinfo {volume} {117}},\ \bibinfo {pages} {087402} (\bibinfo
  {year} {2016})}\BibitemShut {NoStop}%
\bibitem [{\citenamefont {Lim}\ and\ \citenamefont {Moessner}(2017)}]{Lim2017}%
  \BibitemOpen
  \bibfield  {author} {\bibinfo {author} {\bibfnamefont {L.-K.}\ \bibnamefont
  {Lim}}\ and\ \bibinfo {author} {\bibfnamefont {R.}~\bibnamefont {Moessner}},\
  }\href {\doibase 10.1103/PhysRevLett.118.016401} {\bibfield  {journal}
  {\bibinfo  {journal} {Phys. Rev. Lett.}\ }\textbf {\bibinfo {volume} {118}},\
  \bibinfo {pages} {016401} (\bibinfo {year} {2017})}\BibitemShut {NoStop}%
\bibitem [{\citenamefont {Li}\ \emph {et~al.}(2017{\natexlab{a}})\citenamefont
  {Li}, \citenamefont {Yin}, \citenamefont {Chen},\ and\ \citenamefont
  {Ara{\'{u}}jo}}]{Li2017}%
  \BibitemOpen
  \bibfield  {author} {\bibinfo {author} {\bibfnamefont {L.}~\bibnamefont
  {Li}}, \bibinfo {author} {\bibfnamefont {C.}~\bibnamefont {Yin}}, \bibinfo
  {author} {\bibfnamefont {S.}~\bibnamefont {Chen}}, \ and\ \bibinfo {author}
  {\bibfnamefont {M.~A.~N.}\ \bibnamefont {Ara{\'{u}}jo}},\ }\href {\doibase
  10.1103/physrevb.95.121107} {\bibfield  {journal} {\bibinfo  {journal} {Phys.
  Rev. B}\ }\textbf {\bibinfo {volume} {95}},\ \bibinfo {pages} {121107}
  (\bibinfo {year} {2017}{\natexlab{a}})}\BibitemShut {NoStop}%
\bibitem [{\citenamefont {Li}\ \emph {et~al.}(2017{\natexlab{b}})\citenamefont
  {Li}, \citenamefont {Chesi}, \citenamefont {Yin},\ and\ \citenamefont
  {Chen}}]{Li2017_2}%
  \BibitemOpen
  \bibfield  {author} {\bibinfo {author} {\bibfnamefont {L.}~\bibnamefont
  {Li}}, \bibinfo {author} {\bibfnamefont {S.}~\bibnamefont {Chesi}}, \bibinfo
  {author} {\bibfnamefont {C.}~\bibnamefont {Yin}}, \ and\ \bibinfo {author}
  {\bibfnamefont {S.}~\bibnamefont {Chen}},\ }\href {\doibase
  10.1103/PhysRevB.96.081116} {\bibfield  {journal} {\bibinfo  {journal} {Phys.
  Rev. B}\ }\textbf {\bibinfo {volume} {96}},\ \bibinfo {pages} {081116}
  (\bibinfo {year} {2017}{\natexlab{b}})}\BibitemShut {NoStop}%
\bibitem [{\citenamefont {Wu}\ \emph {et~al.}(2016)\citenamefont {Wu},
  \citenamefont {Wang}, \citenamefont {Mun}, \citenamefont {Johnson},
  \citenamefont {Mou}, \citenamefont {Huang}, \citenamefont {Lee},
  \citenamefont {Bud/'ko}, \citenamefont {Canfield},\ and\ \citenamefont
  {Kaminski}}]{Wu2016}%
  \BibitemOpen
  \bibfield  {author} {\bibinfo {author} {\bibfnamefont {Y.}~\bibnamefont
  {Wu}}, \bibinfo {author} {\bibfnamefont {L.-L.}\ \bibnamefont {Wang}},
  \bibinfo {author} {\bibfnamefont {E.}~\bibnamefont {Mun}}, \bibinfo {author}
  {\bibfnamefont {D.~D.}\ \bibnamefont {Johnson}}, \bibinfo {author}
  {\bibfnamefont {D.}~\bibnamefont {Mou}}, \bibinfo {author} {\bibfnamefont
  {L.}~\bibnamefont {Huang}}, \bibinfo {author} {\bibfnamefont
  {Y.}~\bibnamefont {Lee}}, \bibinfo {author} {\bibfnamefont {S.~L.}\
  \bibnamefont {Bud/'ko}}, \bibinfo {author} {\bibfnamefont {P.~C.}\
  \bibnamefont {Canfield}}, \ and\ \bibinfo {author} {\bibfnamefont
  {A.}~\bibnamefont {Kaminski}},\ }\href {\doibase 10.1038/nphys3712}
  {\bibfield  {journal} {\bibinfo  {journal} {Nat Phys}\ }\textbf {\bibinfo
  {volume} {7}},\ \bibinfo {pages} {667} (\bibinfo {year} {2016})}\BibitemShut
  {NoStop}%
\bibitem [{\citenamefont {Bian}\ \emph {et~al.}(2016)\citenamefont {Bian},
  \citenamefont {Chang}, \citenamefont {Sankar}, \citenamefont {Xu},
  \citenamefont {Zheng}, \citenamefont {Neupert}, \citenamefont {Chiu},
  \citenamefont {Huang}, \citenamefont {Chang}, \citenamefont {Belopolski},
  \citenamefont {Sanchez}, \citenamefont {Neupane}, \citenamefont {Alidoust},
  \citenamefont {Liu}, \citenamefont {Wang}, \citenamefont {Lee}, \citenamefont
  {Jeng}, \citenamefont {Zhang}, \citenamefont {Yuan}, \citenamefont {Jia},
  \citenamefont {Bansil}, \citenamefont {Chou}, \citenamefont {Lin},\ and\
  \citenamefont {Hasan}}]{Bian2016}%
  \BibitemOpen
  \bibfield  {author} {\bibinfo {author} {\bibfnamefont {G.}~\bibnamefont
  {Bian}}, \bibinfo {author} {\bibfnamefont {T.-R.}\ \bibnamefont {Chang}},
  \bibinfo {author} {\bibfnamefont {R.}~\bibnamefont {Sankar}}, \bibinfo
  {author} {\bibfnamefont {S.-Y.}\ \bibnamefont {Xu}}, \bibinfo {author}
  {\bibfnamefont {H.}~\bibnamefont {Zheng}}, \bibinfo {author} {\bibfnamefont
  {T.}~\bibnamefont {Neupert}}, \bibinfo {author} {\bibfnamefont {C.-K.}\
  \bibnamefont {Chiu}}, \bibinfo {author} {\bibfnamefont {S.-M.}\ \bibnamefont
  {Huang}}, \bibinfo {author} {\bibfnamefont {G.}~\bibnamefont {Chang}},
  \bibinfo {author} {\bibfnamefont {I.}~\bibnamefont {Belopolski}}, \bibinfo
  {author} {\bibfnamefont {D.~S.}\ \bibnamefont {Sanchez}}, \bibinfo {author}
  {\bibfnamefont {M.}~\bibnamefont {Neupane}}, \bibinfo {author} {\bibfnamefont
  {N.}~\bibnamefont {Alidoust}}, \bibinfo {author} {\bibfnamefont
  {C.}~\bibnamefont {Liu}}, \bibinfo {author} {\bibfnamefont {B.}~\bibnamefont
  {Wang}}, \bibinfo {author} {\bibfnamefont {C.-C.}\ \bibnamefont {Lee}},
  \bibinfo {author} {\bibfnamefont {H.-T.}\ \bibnamefont {Jeng}}, \bibinfo
  {author} {\bibfnamefont {C.}~\bibnamefont {Zhang}}, \bibinfo {author}
  {\bibfnamefont {Z.}~\bibnamefont {Yuan}}, \bibinfo {author} {\bibfnamefont
  {S.}~\bibnamefont {Jia}}, \bibinfo {author} {\bibfnamefont {A.}~\bibnamefont
  {Bansil}}, \bibinfo {author} {\bibfnamefont {F.}~\bibnamefont {Chou}},
  \bibinfo {author} {\bibfnamefont {H.}~\bibnamefont {Lin}}, \ and\ \bibinfo
  {author} {\bibfnamefont {M.~Z.}\ \bibnamefont {Hasan}},\ }\href {\doibase
  10.1038/ncomms10556} {\bibfield  {journal} {\bibinfo  {journal} {Nat. Comm.}\
  }\textbf {\bibinfo {volume} {7}},\ \bibinfo {pages} {10556} (\bibinfo {year}
  {2016})}\BibitemShut {NoStop}%
\bibitem [{\citenamefont {Hu}\ \emph {et~al.}(2016)\citenamefont {Hu},
  \citenamefont {Tang}, \citenamefont {Liu}, \citenamefont {Liu}, \citenamefont
  {Zhu}, \citenamefont {Graf}, \citenamefont {Myhro}, \citenamefont {Tran},
  \citenamefont {Lau}, \citenamefont {Wei},\ and\ \citenamefont
  {Mao}}]{Hu2016}%
  \BibitemOpen
  \bibfield  {author} {\bibinfo {author} {\bibfnamefont {J.}~\bibnamefont
  {Hu}}, \bibinfo {author} {\bibfnamefont {Z.}~\bibnamefont {Tang}}, \bibinfo
  {author} {\bibfnamefont {J.}~\bibnamefont {Liu}}, \bibinfo {author}
  {\bibfnamefont {X.}~\bibnamefont {Liu}}, \bibinfo {author} {\bibfnamefont
  {Y.}~\bibnamefont {Zhu}}, \bibinfo {author} {\bibfnamefont {D.}~\bibnamefont
  {Graf}}, \bibinfo {author} {\bibfnamefont {K.}~\bibnamefont {Myhro}},
  \bibinfo {author} {\bibfnamefont {S.}~\bibnamefont {Tran}}, \bibinfo {author}
  {\bibfnamefont {C.~N.}\ \bibnamefont {Lau}}, \bibinfo {author} {\bibfnamefont
  {J.}~\bibnamefont {Wei}}, \ and\ \bibinfo {author} {\bibfnamefont
  {Z.}~\bibnamefont {Mao}},\ }\href {\doibase 10.1103/PhysRevLett.117.016602}
  {\bibfield  {journal} {\bibinfo  {journal} {Phys. Rev. Lett.}\ }\textbf
  {\bibinfo {volume} {117}},\ \bibinfo {pages} {016602} (\bibinfo {year}
  {2016})}\BibitemShut {NoStop}%
\bibitem [{\citenamefont {Yan}\ \emph {et~al.}(2017{\natexlab{a}})\citenamefont
  {Yan}, \citenamefont {Liu}, \citenamefont {Yan}, \citenamefont {Liu},
  \citenamefont {Chen}, \citenamefont {Wang},\ and\ \citenamefont
  {Lu}}]{Yan2017}%
  \BibitemOpen
  \bibfield  {author} {\bibinfo {author} {\bibfnamefont {Q.}~\bibnamefont
  {Yan}}, \bibinfo {author} {\bibfnamefont {R.}~\bibnamefont {Liu}}, \bibinfo
  {author} {\bibfnamefont {Z.}~\bibnamefont {Yan}}, \bibinfo {author}
  {\bibfnamefont {B.}~\bibnamefont {Liu}}, \bibinfo {author} {\bibfnamefont
  {H.}~\bibnamefont {Chen}}, \bibinfo {author} {\bibfnamefont {Z.}~\bibnamefont
  {Wang}}, \ and\ \bibinfo {author} {\bibfnamefont {L.}~\bibnamefont {Lu}},\
  }\href@noop {} {\bibfield  {journal} {\bibinfo  {journal} {arXiv}\ }
  (\bibinfo {year} {2017}{\natexlab{a}})},\ \Eprint
  {http://arxiv.org/abs/1706.05500v1} {1706.05500v1} \BibitemShut {NoStop}%
\bibitem [{\citenamefont {Wan}\ \emph {et~al.}(2011)\citenamefont {Wan},
  \citenamefont {Turner}, \citenamefont {Vishwanath},\ and\ \citenamefont
  {Savrasov}}]{Wan2011}%
  \BibitemOpen
  \bibfield  {author} {\bibinfo {author} {\bibfnamefont {X.}~\bibnamefont
  {Wan}}, \bibinfo {author} {\bibfnamefont {A.~M.}\ \bibnamefont {Turner}},
  \bibinfo {author} {\bibfnamefont {A.}~\bibnamefont {Vishwanath}}, \ and\
  \bibinfo {author} {\bibfnamefont {S.~Y.}\ \bibnamefont {Savrasov}},\
  }\href@noop {} {\bibfield  {journal} {\bibinfo  {journal} {Phys. Rev. B}\
  }\textbf {\bibinfo {volume} {83}},\ \bibinfo {pages} {205101} (\bibinfo
  {year} {2011})}\BibitemShut {NoStop}%
\bibitem [{\citenamefont {Young}\ \emph {et~al.}(2012)\citenamefont {Young},
  \citenamefont {Zaheer}, \citenamefont {Teo}, \citenamefont {Kane},
  \citenamefont {Mele},\ and\ \citenamefont {Rappe}}]{Young2012}%
  \BibitemOpen
  \bibfield  {author} {\bibinfo {author} {\bibfnamefont {S.~M.}\ \bibnamefont
  {Young}}, \bibinfo {author} {\bibfnamefont {S.}~\bibnamefont {Zaheer}},
  \bibinfo {author} {\bibfnamefont {J.~C.~Y.}\ \bibnamefont {Teo}}, \bibinfo
  {author} {\bibfnamefont {C.~L.}\ \bibnamefont {Kane}}, \bibinfo {author}
  {\bibfnamefont {E.~J.}\ \bibnamefont {Mele}}, \ and\ \bibinfo {author}
  {\bibfnamefont {A.~M.}\ \bibnamefont {Rappe}},\ }\href@noop {} {\bibfield
  {journal} {\bibinfo  {journal} {Phys. Rev. Lett.}\ }\textbf {\bibinfo
  {volume} {108}},\ \bibinfo {pages} {140405} (\bibinfo {year}
  {2012})}\BibitemShut {NoStop}%
\bibitem [{\citenamefont {Morimoto}\ and\ \citenamefont
  {Furusaki}(2014)}]{Morimoto2014}%
  \BibitemOpen
  \bibfield  {author} {\bibinfo {author} {\bibfnamefont {T.}~\bibnamefont
  {Morimoto}}\ and\ \bibinfo {author} {\bibfnamefont {A.}~\bibnamefont
  {Furusaki}},\ }\href@noop {} {\bibfield  {journal} {\bibinfo  {journal}
  {Phys. Rev. B}\ }\textbf {\bibinfo {volume} {89}},\ \bibinfo {pages} {235127}
  (\bibinfo {year} {2014})}\BibitemShut {NoStop}%
\bibitem [{\citenamefont {Yang}\ and\ \citenamefont
  {Nagaosa}(2014)}]{Yang2014}%
  \BibitemOpen
  \bibfield  {author} {\bibinfo {author} {\bibfnamefont {B.-J.}\ \bibnamefont
  {Yang}}\ and\ \bibinfo {author} {\bibfnamefont {N.}~\bibnamefont {Nagaosa}},\
  }\href {\doibase 10.1038/ncomms5898} {\bibfield  {journal} {\bibinfo
  {journal} {Nat. Commun.}\ }\textbf {\bibinfo {volume} {5}},\ \bibinfo {pages}
  {4898} (\bibinfo {year} {2014})}\BibitemShut {NoStop}%
\bibitem [{\citenamefont {Zhong}\ \emph {et~al.}(2017)\citenamefont {Zhong},
  \citenamefont {Chen}, \citenamefont {Yu}, \citenamefont {Xie}, \citenamefont
  {Wang}, \citenamefont {Yang},\ and\ \citenamefont {Zhang}}]{Zhong2017}%
  \BibitemOpen
  \bibfield  {author} {\bibinfo {author} {\bibfnamefont {C.}~\bibnamefont
  {Zhong}}, \bibinfo {author} {\bibfnamefont {Y.}~\bibnamefont {Chen}},
  \bibinfo {author} {\bibfnamefont {Z.-M.}\ \bibnamefont {Yu}}, \bibinfo
  {author} {\bibfnamefont {Y.}~\bibnamefont {Xie}}, \bibinfo {author}
  {\bibfnamefont {H.}~\bibnamefont {Wang}}, \bibinfo {author} {\bibfnamefont
  {S.~A.}\ \bibnamefont {Yang}}, \ and\ \bibinfo {author} {\bibfnamefont
  {S.}~\bibnamefont {Zhang}},\ }\href {http://dx.doi.org/10.1038/ncomms15641}
  {\bibfield  {journal} {\bibinfo  {journal} {Nat. Comm.}\ }\textbf {\bibinfo
  {volume} {8}},\ \bibinfo {pages} {15641} (\bibinfo {year}
  {2017})}\BibitemShut {NoStop}%
\bibitem [{\citenamefont {Chen}\ \emph {et~al.}(2017)\citenamefont {Chen},
  \citenamefont {Lu},\ and\ \citenamefont {Hou}}]{Chen_Hopf}%
  \BibitemOpen
  \bibfield  {author} {\bibinfo {author} {\bibfnamefont {W.}~\bibnamefont
  {Chen}}, \bibinfo {author} {\bibfnamefont {H.-Z.}\ \bibnamefont {Lu}}, \ and\
  \bibinfo {author} {\bibfnamefont {J.-M.}\ \bibnamefont {Hou}},\ }\href
  {\doibase 10.1103/PhysRevB.96.041102} {\bibfield  {journal} {\bibinfo
  {journal} {Phys. Rev. B}\ }\textbf {\bibinfo {volume} {96}},\ \bibinfo
  {pages} {041102} (\bibinfo {year} {2017})}\BibitemShut {NoStop}%
\bibitem [{\citenamefont {Yan}\ \emph {et~al.}(2017{\natexlab{b}})\citenamefont
  {Yan}, \citenamefont {Bi}, \citenamefont {Shen}, \citenamefont {Lu},
  \citenamefont {Zhang},\ and\ \citenamefont {Wang}}]{Yan_Hopf}%
  \BibitemOpen
  \bibfield  {author} {\bibinfo {author} {\bibfnamefont {Z.}~\bibnamefont
  {Yan}}, \bibinfo {author} {\bibfnamefont {R.}~\bibnamefont {Bi}}, \bibinfo
  {author} {\bibfnamefont {H.}~\bibnamefont {Shen}}, \bibinfo {author}
  {\bibfnamefont {L.}~\bibnamefont {Lu}}, \bibinfo {author} {\bibfnamefont
  {S.-C.}\ \bibnamefont {Zhang}}, \ and\ \bibinfo {author} {\bibfnamefont
  {Z.}~\bibnamefont {Wang}},\ }\href {\doibase 10.1103/PhysRevB.96.041103}
  {\bibfield  {journal} {\bibinfo  {journal} {Phys. Rev. B}\ }\textbf {\bibinfo
  {volume} {96}},\ \bibinfo {pages} {041103} (\bibinfo {year}
  {2017}{\natexlab{b}})}\BibitemShut {NoStop}%
\bibitem [{\citenamefont {Ezawa}(2017{\natexlab{a}})}]{Ezawa_Hopf}%
  \BibitemOpen
  \bibfield  {author} {\bibinfo {author} {\bibfnamefont {M.}~\bibnamefont
  {Ezawa}},\ }\href {\doibase 10.1103/PhysRevB.96.041202} {\bibfield  {journal}
  {\bibinfo  {journal} {Phys. Rev. B}\ }\textbf {\bibinfo {volume} {96}},\
  \bibinfo {pages} {041202} (\bibinfo {year} {2017}{\natexlab{a}})}\BibitemShut
  {NoStop}%
\bibitem [{\citenamefont {Chang}\ and\ \citenamefont {Yee}(2017)}]{Chang_Hopf}%
  \BibitemOpen
  \bibfield  {author} {\bibinfo {author} {\bibfnamefont {P.-Y.}\ \bibnamefont
  {Chang}}\ and\ \bibinfo {author} {\bibfnamefont {C.-H.}\ \bibnamefont
  {Yee}},\ }\href {\doibase 10.1103/PhysRevB.96.081114} {\bibfield  {journal}
  {\bibinfo  {journal} {Phys. Rev. B}\ }\textbf {\bibinfo {volume} {96}},\
  \bibinfo {pages} {081114} (\bibinfo {year} {2017})}\BibitemShut {NoStop}%
\bibitem [{\citenamefont {Bi}\ \emph {et~al.}()\citenamefont {Bi},
  \citenamefont {Yan}, \citenamefont {Lu},\ and\ \citenamefont
  {Wang}}]{Bi2017}%
  \BibitemOpen
  \bibfield  {author} {\bibinfo {author} {\bibfnamefont {R.}~\bibnamefont
  {Bi}}, \bibinfo {author} {\bibfnamefont {Z.}~\bibnamefont {Yan}}, \bibinfo
  {author} {\bibfnamefont {L.}~\bibnamefont {Lu}}, \ and\ \bibinfo {author}
  {\bibfnamefont {Z.}~\bibnamefont {Wang}},\ }\href@noop {} {\bibfield
  {journal} {\bibinfo  {journal} {arXiv}\ }}\Eprint
  {http://arxiv.org/abs/1704.06849v2} {1704.06849v2} \BibitemShut {NoStop}%
\bibitem [{\citenamefont {Bzdušek}\ \emph {et~al.}(2016)\citenamefont
  {Bzdušek}, \citenamefont {Wu}, \citenamefont {Rüegg}, \citenamefont
  {Sigrist},\ and\ \citenamefont {Soluyanov}}]{Bzdusek2016}%
  \BibitemOpen
  \bibfield  {author} {\bibinfo {author} {\bibfnamefont {T.}~\bibnamefont
  {Bzdušek}}, \bibinfo {author} {\bibfnamefont {Q.}~\bibnamefont {Wu}},
  \bibinfo {author} {\bibfnamefont {A.}~\bibnamefont {Rüegg}}, \bibinfo
  {author} {\bibfnamefont {M.}~\bibnamefont {Sigrist}}, \ and\ \bibinfo
  {author} {\bibfnamefont {A.~A.}\ \bibnamefont {Soluyanov}},\ }\href {\doibase
  10.1038/nature19099} {\bibfield  {journal} {\bibinfo  {journal} {Nature}\
  }\textbf {\bibinfo {volume} {538}},\ \bibinfo {pages} {75} (\bibinfo {year}
  {2016})}\BibitemShut {NoStop}%
\bibitem [{\citenamefont {Yan}\ and\ \citenamefont
  {Wang}(2017)}]{Yan_crossing}%
  \BibitemOpen
  \bibfield  {author} {\bibinfo {author} {\bibfnamefont {Z.}~\bibnamefont
  {Yan}}\ and\ \bibinfo {author} {\bibfnamefont {Z.}~\bibnamefont {Wang}},\
  }\href {\doibase 10.1103/PhysRevB.96.041206} {\bibfield  {journal} {\bibinfo
  {journal} {Phys. Rev. B}\ }\textbf {\bibinfo {volume} {96}},\ \bibinfo
  {pages} {041206} (\bibinfo {year} {2017})}\BibitemShut {NoStop}%
\bibitem [{\citenamefont {Ezawa}(2017{\natexlab{b}})}]{Ezawa_crossing}%
  \BibitemOpen
  \bibfield  {author} {\bibinfo {author} {\bibfnamefont {M.}~\bibnamefont
  {Ezawa}},\ }\href {\doibase 10.1103/PhysRevB.96.041205} {\bibfield  {journal}
  {\bibinfo  {journal} {Phys. Rev. B}\ }\textbf {\bibinfo {volume} {96}},\
  \bibinfo {pages} {041205} (\bibinfo {year} {2017}{\natexlab{b}})}\BibitemShut
  {NoStop}%
\bibitem [{\citenamefont {Okugawa}\ and\ \citenamefont
  {Murakami}(2017)}]{Okugawa2017}%
  \BibitemOpen
  \bibfield  {author} {\bibinfo {author} {\bibfnamefont {R.}~\bibnamefont
  {Okugawa}}\ and\ \bibinfo {author} {\bibfnamefont {S.}~\bibnamefont
  {Murakami}},\ }\href {\doibase 10.1103/PhysRevB.96.115201} {\bibfield
  {journal} {\bibinfo  {journal} {Phys. Rev. B}\ }\textbf {\bibinfo {volume}
  {96}},\ \bibinfo {pages} {115201} (\bibinfo {year} {2017})}\BibitemShut
  {NoStop}%
\bibitem [{\citenamefont {Lifshitz}(1960)}]{Lifshitz}%
  \BibitemOpen
  \bibfield  {author} {\bibinfo {author} {\bibfnamefont {I.~M.}\ \bibnamefont
  {Lifshitz}},\ }\href@noop {} {\bibfield  {journal} {\bibinfo  {journal} {Sov.
  Phys. JETP}\ }\textbf {\bibinfo {volume} {11}},\ \bibinfo {pages} {1130}
  (\bibinfo {year} {1960})}\BibitemShut {NoStop}%
\bibitem [{\citenamefont {Bouhon}\ and\ \citenamefont
  {Black-Schaffer}()}]{Lifshitz_NL}%
  \BibitemOpen
  \bibfield  {author} {\bibinfo {author} {\bibfnamefont {A.}~\bibnamefont
  {Bouhon}}\ and\ \bibinfo {author} {\bibfnamefont {A.~M.}\ \bibnamefont
  {Black-Schaffer}},\ }\href@noop {} {\bibfield  {journal} {\bibinfo  {journal}
  {arXiv}\ }}\Eprint {http://arxiv.org/abs/1710.04871v1} {1710.04871v1}
  \BibitemShut {NoStop}%
\bibitem [{\citenamefont {Li}\ and\ \citenamefont
  {Ara{\'{u}}jo}(2016)}]{Li2016}%
  \BibitemOpen
  \bibfield  {author} {\bibinfo {author} {\bibfnamefont {L.}~\bibnamefont
  {Li}}\ and\ \bibinfo {author} {\bibfnamefont {M.~A.~N.}\ \bibnamefont
  {Ara{\'{u}}jo}},\ }\href {\doibase 10.1103/physrevb.94.165117} {\bibfield
  {journal} {\bibinfo  {journal} {Phys. Rev. B}\ }\textbf {\bibinfo {volume}
  {94}},\ \bibinfo {pages} {165117} (\bibinfo {year} {2016})}\BibitemShut
  {NoStop}%
\bibitem [{\citenamefont {Türker}\ and\ \citenamefont {Moroz}()}]{Turker2017}%
  \BibitemOpen
  \bibfield  {author} {\bibinfo {author} {\bibfnamefont {O.}~\bibnamefont
  {Türker}}\ and\ \bibinfo {author} {\bibfnamefont {S.}~\bibnamefont
  {Moroz}},\ }\href@noop {} {\bibfield  {journal} {\bibinfo  {journal} {arXiv}\
  }}\Eprint {http://arxiv.org/abs/1709.01561v1} {1709.01561v1} \BibitemShut
  {NoStop}%
\bibitem [{\citenamefont {Li}\ \emph {et~al.}(2017{\natexlab{c}})\citenamefont
  {Li}, \citenamefont {Yap}, \citenamefont {Ara{\'{u}}jo},\ and\ \citenamefont
  {Gong}}]{Li2017_3}%
  \BibitemOpen
  \bibfield  {author} {\bibinfo {author} {\bibfnamefont {L.}~\bibnamefont
  {Li}}, \bibinfo {author} {\bibfnamefont {H.~H.}\ \bibnamefont {Yap}},
  \bibinfo {author} {\bibfnamefont {M.~A.~N.}\ \bibnamefont {Ara{\'{u}}jo}}, \
  and\ \bibinfo {author} {\bibfnamefont {J.}~\bibnamefont {Gong}},\ }\href@noop
  {} {\bibfield  {journal} {\bibinfo  {journal} {arXiv}\ } (\bibinfo {year}
  {2017}{\natexlab{c}})},\ \Eprint {http://arxiv.org/abs/1709.07132v1}
  {1709.07132v1} \BibitemShut {NoStop}%
\bibitem [{\citenamefont {Albert}\ \emph {et~al.}(2015)\citenamefont {Albert},
  \citenamefont {Glazman},\ and\ \citenamefont {Jiang}}]{circuit1}%
  \BibitemOpen
  \bibfield  {author} {\bibinfo {author} {\bibfnamefont {V.~V.}\ \bibnamefont
  {Albert}}, \bibinfo {author} {\bibfnamefont {L.~I.}\ \bibnamefont {Glazman}},
  \ and\ \bibinfo {author} {\bibfnamefont {L.}~\bibnamefont {Jiang}},\ }\href
  {\doibase 10.1103/PhysRevLett.114.173902} {\bibfield  {journal} {\bibinfo
  {journal} {Phys. Rev. Lett.}\ }\textbf {\bibinfo {volume} {114}},\ \bibinfo
  {pages} {173902} (\bibinfo {year} {2015})}\BibitemShut {NoStop}%
\bibitem [{\citenamefont {Ningyuan}\ \emph {et~al.}(2015)\citenamefont
  {Ningyuan}, \citenamefont {Owens}, \citenamefont {Sommer}, \citenamefont
  {Schuster},\ and\ \citenamefont {Simon}}]{circuit2}%
  \BibitemOpen
  \bibfield  {author} {\bibinfo {author} {\bibfnamefont {J.}~\bibnamefont
  {Ningyuan}}, \bibinfo {author} {\bibfnamefont {C.}~\bibnamefont {Owens}},
  \bibinfo {author} {\bibfnamefont {A.}~\bibnamefont {Sommer}}, \bibinfo
  {author} {\bibfnamefont {D.}~\bibnamefont {Schuster}}, \ and\ \bibinfo
  {author} {\bibfnamefont {J.}~\bibnamefont {Simon}},\ }\href {\doibase
  10.1103/PhysRevX.5.021031} {\bibfield  {journal} {\bibinfo  {journal} {Phys.
  Rev. X}\ }\textbf {\bibinfo {volume} {5}},\ \bibinfo {pages} {021031}
  (\bibinfo {year} {2015})}\BibitemShut {NoStop}%
\bibitem [{\citenamefont {Lee}\ \emph {et~al.}()\citenamefont {Lee},
  \citenamefont {Imhof}, \citenamefont {Berger}, \citenamefont {Bayer},
  \citenamefont {Brehm}, \citenamefont {Molenkamp}, \citenamefont {Kiessling},\
  and\ \citenamefont {Thomale}}]{circuit3}%
  \BibitemOpen
  \bibfield  {author} {\bibinfo {author} {\bibfnamefont {C.~H.}\ \bibnamefont
  {Lee}}, \bibinfo {author} {\bibfnamefont {S.}~\bibnamefont {Imhof}}, \bibinfo
  {author} {\bibfnamefont {C.}~\bibnamefont {Berger}}, \bibinfo {author}
  {\bibfnamefont {F.}~\bibnamefont {Bayer}}, \bibinfo {author} {\bibfnamefont
  {J.}~\bibnamefont {Brehm}}, \bibinfo {author} {\bibfnamefont {L.~W.}\
  \bibnamefont {Molenkamp}}, \bibinfo {author} {\bibfnamefont {T.}~\bibnamefont
  {Kiessling}}, \ and\ \bibinfo {author} {\bibfnamefont {R.}~\bibnamefont
  {Thomale}},\ }\href@noop {} {\bibfield  {journal} {\bibinfo  {journal}
  {arXiv}\ }}\Eprint {http://arxiv.org/abs/1705.01077v3} {1705.01077v3}
  \BibitemShut {NoStop}%
\bibitem [{\citenamefont {Imhof}\ \emph {et~al.}()\citenamefont {Imhof},
  \citenamefont {Berger}, \citenamefont {Bayer}, \citenamefont {Brehm},
  \citenamefont {Molenkamp}, \citenamefont {Kiessling}, \citenamefont
  {Schindler}, \citenamefont {Lee}, \citenamefont {Greiter}, \citenamefont
  {Neupert},\ and\ \citenamefont {Thomale}}]{circuit4}%
  \BibitemOpen
  \bibfield  {author} {\bibinfo {author} {\bibfnamefont {S.}~\bibnamefont
  {Imhof}}, \bibinfo {author} {\bibfnamefont {C.}~\bibnamefont {Berger}},
  \bibinfo {author} {\bibfnamefont {F.}~\bibnamefont {Bayer}}, \bibinfo
  {author} {\bibfnamefont {J.}~\bibnamefont {Brehm}}, \bibinfo {author}
  {\bibfnamefont {L.}~\bibnamefont {Molenkamp}}, \bibinfo {author}
  {\bibfnamefont {T.}~\bibnamefont {Kiessling}}, \bibinfo {author}
  {\bibfnamefont {F.}~\bibnamefont {Schindler}}, \bibinfo {author}
  {\bibfnamefont {C.~H.}\ \bibnamefont {Lee}}, \bibinfo {author} {\bibfnamefont
  {M.}~\bibnamefont {Greiter}}, \bibinfo {author} {\bibfnamefont
  {T.}~\bibnamefont {Neupert}}, \ and\ \bibinfo {author} {\bibfnamefont
  {R.}~\bibnamefont {Thomale}},\ }\href@noop {} {\bibfield  {journal} {\bibinfo
   {journal} {arXiv}\ }}\Eprint {http://arxiv.org/abs/1708.03647v1}
  {1708.03647v1} \BibitemShut {NoStop}%
\end{thebibliography}%


\end{document}